\documentclass[runningheads]{llncs}
\usepackage{graphicx,url,hyperref}

\title{An Initial study on Birdsong Re-synthesis Using Neural Vocoders}

\begin{document}
\author{Rhythm Bhatia \and
Tomi H. Kinnunen}
\institute{University of Eastern Finland \\ \email{\{rhythm.bhatia,tkinnu\}@cs.uef.fi}\\}

\maketitle
\begin{abstract}
Modern speech synthesis uses neural vocoders to model raw waveform samples directly. This increased versatility has expanded the scope of vocoders from speech to other domains, such as music. We address another interesting domain of bio-acoustics. We provide initial comparative analysis-resynthesis experiments of birdsong using traditional (WORLD) and two neural (WaveNet autoencoder, parallel WaveGAN) vocoders. Our subjective results indicate no difference in the three vocoders in terms of species discrimination (ABX test). Nonetheless, the WORLD vocoder samples were rated higher in terms of retaining bird-like qualities (MOS test). All  vocoders faced issues with pitch and voicing. Our results  indicate some of the challenges in processing low-quality wildlife audio data.
\end{abstract}
\noindent\textbf{Index Terms}: bioacoustics, neural vocoding, birdsong

\section{Introduction}

\label{sec:intro}

Birdsong is a subject of intensive research. Richness and variety of birdsong has intrigued basic research into the communicative function of birdsong. Besides ecology, birdsong has raised interest within the signal processing and machine learning communities too. This includes tasks such as species recognition \cite{Somervuo2006-parametric-bird} and locating active bird segments (akin to speech activity detection) \cite{Stowell2016-bird}. 

While a number of recognition, segmentation and labeling approaches has been proposed, \emph{generation} of birdsong has received less attention. There are, however, many potential applications ranging from games, movies and virtual reality to education and robotics where flexible generation of birdsong (and other animal vocalizations) could be useful. 

Most prior work considers specialized physical models (e.g. \cite{Moore16-mammalian-synthesizer,Amador2021-synthetic}) or adapts speech vocoders to birdsong (e.g. \cite{OReilly2016-YIN-Bird}). A limited number of studies also use \emph{text-to-speech} (TTS) techniques to synthesize birdsong \cite{Bonada2016-HMM-synthesis,Gutscher2019-budgerigar}. 
An obvious difficulty is that, even if birdsong and speech both serve a communicative function, `bird language' lacks a commonly-agreed, standard written form (ortography). Human language exists both in written and spoken forms and the statistical association between the two enable tasks such as TTS or automatic speech recognition. In the limited number of `bird TTS' studies, the problem has been addressed using acoustic units learnt using unsupervised techniques. Similar approaches have recently been addressed by the speech community \cite{dunbar2019zero,tjandra2019vqvae}. 

We address birdsong generation using \textbf{neural waveform models}.  Traditional TTS methods use fixed (signal processing based) operations to represent speech waveforms using a small number of parameters --- such as spectral envelope, fundamental frequency and aperiodicity. A major breakthrough was brought in 2016 by the introduction of \emph{WaveNet}  \cite{VandenOord2016-wavenet}, an approach to model raw waveform samples directly. WaveNet and other neural waveform models have provided excellent results in modeling other acoustic signals beyond speech --- such as music. However, unlike physical models and traditional vocoders, the neural models require (often time-consuming) training and, similar to any machine learning task, can be sensitive to choice of training data, architectures and control parameters.

As far as the authors are aware, our work is the first to address birdsong generation using neural waveform models.
We purposefully limit our focus to the \textbf{vocoder} part only, which is a key component of complete synthesizers. Given limited work in this domain, we feel the selected focus is justified. To this end, we have chosen three modern vocoders --- WORLD \cite{Morise2016-WORLD}, WaveNet Autoencoder \cite{engel2017neural} and Parallel WaveGAN \cite{yamamoto2020parallel}. We compare them through analysis-resynthesis experiments, including objective and subjective evaluation.

\section{Selected Vocoders}
\label{sec:pagestyle}

Our aim is to compare traditional and neural vocoders in their ability to re-synthesize birdsong. We consider three popular vocoders using architectures and parameter settings from the original code repositories (\href{https://github.com/JeremyCCHsu/Python-Wrapper-for-World-Vocoder}{WORLD}, \href{https://github.com/magenta/magenta/tree/master/magenta/models/nsynth}{WaveNet AE}, \href{https://github.com/kan-bayashi/ParallelWaveGAN} {Parallel WaveGAN}). The neural vocoders are retrained on birdsong (detailed below) without further parameter tuning. While the main reason is computational, we also want to find out how well the architectures designed for speech cope with birdsong `off-the-shelf'.

\textbf{WORLD} \cite{Morise2016-WORLD} is a signal processing based vocoder without trainable components. It decomposes a waveform into fundamental frequency (F0), spectral envelope, and aperiodicity parameters. Being an alternative to classic vocoders such as STRAIGHT \cite{Kawahara1999-STRAIGHT} and TANDEM-STRAIGHT \cite{Kawahara2008-tandem-STRAIGHT}, WORLD is adopted in many text-to-speech and voice conversion studies. We use D4C edition \cite{morise2016d4c} of WORLD.

\textbf{WaveNet autoencoder} \cite{engel2017neural} is a data-driven model that learns temporal hidden codes from training data and models long-term information using an \emph{encoder-decoder} structure. The encoder infers hidden embeddings distributed in time which the decoder uses to reconstruct the input. The temporal encoder has the same dilation block as WaveNet but uses \emph{non-causal} convolutions by considering the entire input context. The model consists of 30 convolutional layers followed by an average pooling layer to create a temporal embedding of $16$ dimensions every $512$ samples. The vanilla WaveNet decoder with $30$ layers (each layer being $1 \times 1$ convolution along with a bias) is used to upsample the embedding back to the original time resolution. The model is trained for $100$k iterations with a batch size of $32$.

\textbf{Parallel WaveGAN} \cite{yamamoto2020parallel} is a small footprint waveform generation method based on \emph{generative adversarial network} (GAN) \cite{Goodfellow2014-GAN}. It is trained jointly on multi resolution short-time Fourier transform loss and waveform domain adverserial loss. The model consists of $30$ convolutionally dilated layers with exponentially increasing $3$ dilation cycles with $64$ residual and skip channels and filter size of $3$. The discriminator consists of $10$ non-causal dilated 1-D convolutions with leaky ReLU activation. 
Linearly increasing dilations in the range of one to eight along with the stride of $1$ are applied for the 1D convolutions apart from the first and the last layer. Weight normalization is applied to all the convolutional layers for both the generator and the discriminator \cite{salimans2016weight}. The model is trained for 100k steps using RAdam optimizer. The discriminator was fixed for first 50k steps after which both the models were trained jointly. The minibatch size was set to eight along with the 24k timesamples audioclip i.e. length of each audio clip was 1.0 second. The initial learning rate for generator is 0.0001 and for discriminator 0.00005.

\section{Experimental Set-up}

\subsection{Dataset}
We use a subset of \emph{xccoverbl} dataset \cite{stowell2014automatic} in our experiments. It contains vocalizations of 88 bird species commonly observed in the UK, gathered from the large \emph{Xeno Canto} collection \cite{web:xenocanto}.The sampling rate is $44.1$kHz. For all the 88 species, 3 audio files (each 30 seconds of total audio) are used for training and 2 other files (each 20 seconds) are held out for testing purposes. Our experiments include both objective and subjective evaluation. Given the time-consuming nature of the latter, we limit the experiments reported to 10 bird species in total, selected randomly out from the 88.

\subsection{Objective Evaluation} 
We use \emph{root mean square error} (RMSE) to measure distance of the mel cepstra of the original and resynthesized birdsong as our objective quality measure. The lower the RMSE, the higher the (objective) quality \cite{haque2018conditional}. The MFCCs used in RMSE computation are calculated using the speech signal processing tool kit (SPTK) \cite{imai2009speech}.

\subsection{Subjective evaluation - species discrimination (ABX)}

Our subjective experiments serve to address two questions: how much vocoding influences (1) bird species discrimination; and (2) preservation of bird-related traits. We adopt an ABX test and a quality rating experiment to address these questions, respectively. In practice, we gather the responses to both simultaneously.

For the ABX test, each listener is provided with triplets (trials) of audio files. A and B correspond to natural audio of two different species, and X represents either a natural or a resynthesized sample of one one of the species. The listeners are asked to choose whether X resembles more A or B. We prepare a total of 10 trials (ABX triplets) per each of the three vocoders. Along with the additional case of natural audio (no vocoding) this yields a total of 40 ABX trials per subject.

The A and B samples in a given trial are selected from the training set of the vocoders. X is always selected from a \emph{different} audio file corresponding to species of either A or B; this way, the subject cannot do trivial `content-matched' comparison of original-vs-resynthesized file, but will have to pay attention to the general properties of the two species. We summarize the results of the ABX test as percent-correct identification rate (from pooled listener responses) broken down according to the vocoder. 

In preparing the listening test, the order of the trials is randomized, with different order for each subject. The duration of each of the 40 samples was fixed to $10$ seconds (of total audio). All files prepared for the subjective test were additionally normalized using \texttt{ffmpeg-normalize} tool \cite{web:ffmpeg}.

\subsection{Subjective evaluation - bird-related cues (MOS)}

Evaluation of bird-related cues, in turn, is based on 5-point mean opinion (MOS) score ratings. The subject is asked to rate the X sample in each of the ABX trials in an ordinal scale $[1\dots5]$, where $1$ means that X does not resemble bird sounds at all while $5$ means X resembles perfectly bird sound. 

The samples were presented through a PHP-based web-forms. The subjects were free to listen to the samples as many times as needed with their own pace. The subjects took part voluntarily and were informed about the study aims, with standard consent forms provided by the authors' institution in place. No compensations were provided. 

We recruited an initial pool of $17$ subjects. We excluded afterwards two subjects who obtained, respectively, only 2 and 3 correct responses in the ABX test on the 10 natural samples. The results summarized below correspond to the responses of the remaining 15 subjects, all who obtained at least 5/10 correct on the natural samples. 

\section{Experimental Results}

\subsection{Objective Evaluation Results}

The RMSE results shown in Table \ref{tab:rmse} are computed from the same audio files presented to the listeners in the next subsection (10 audio files per method). The results indicate the lowest and highest values for WORLD and WaveNet autoencoder, respectively, with parallel WaveGAN between the two. The higher RMSE values for the two neural approaches suggest issues in generalization from training set to the test data.

\begin{table}[ht]
\caption{Average root mean square error (RMSE) along with 95\% confidence range from standard error of mean (SEM).}
\center
\begin{tabular}{|c| c|} 
 \hline
 Model & RMSE \\  
 \hline\hline
 WORLD & $0.6879 \pm 0.41$\\ 
 \hline
 WaveNet Autoencoder & $3.2047 \pm 1.20$  \\ 
 \hline
 Parallel WaveGAN & $1.82647 \pm 0.67$ \\ 
 \hline
\end{tabular}\label{tab:rmse}
\end{table}

\subsection{Subjective results: species discrimination (ABX)}

\begin{figure}[t]
  \centering
  \includegraphics[width=\linewidth]{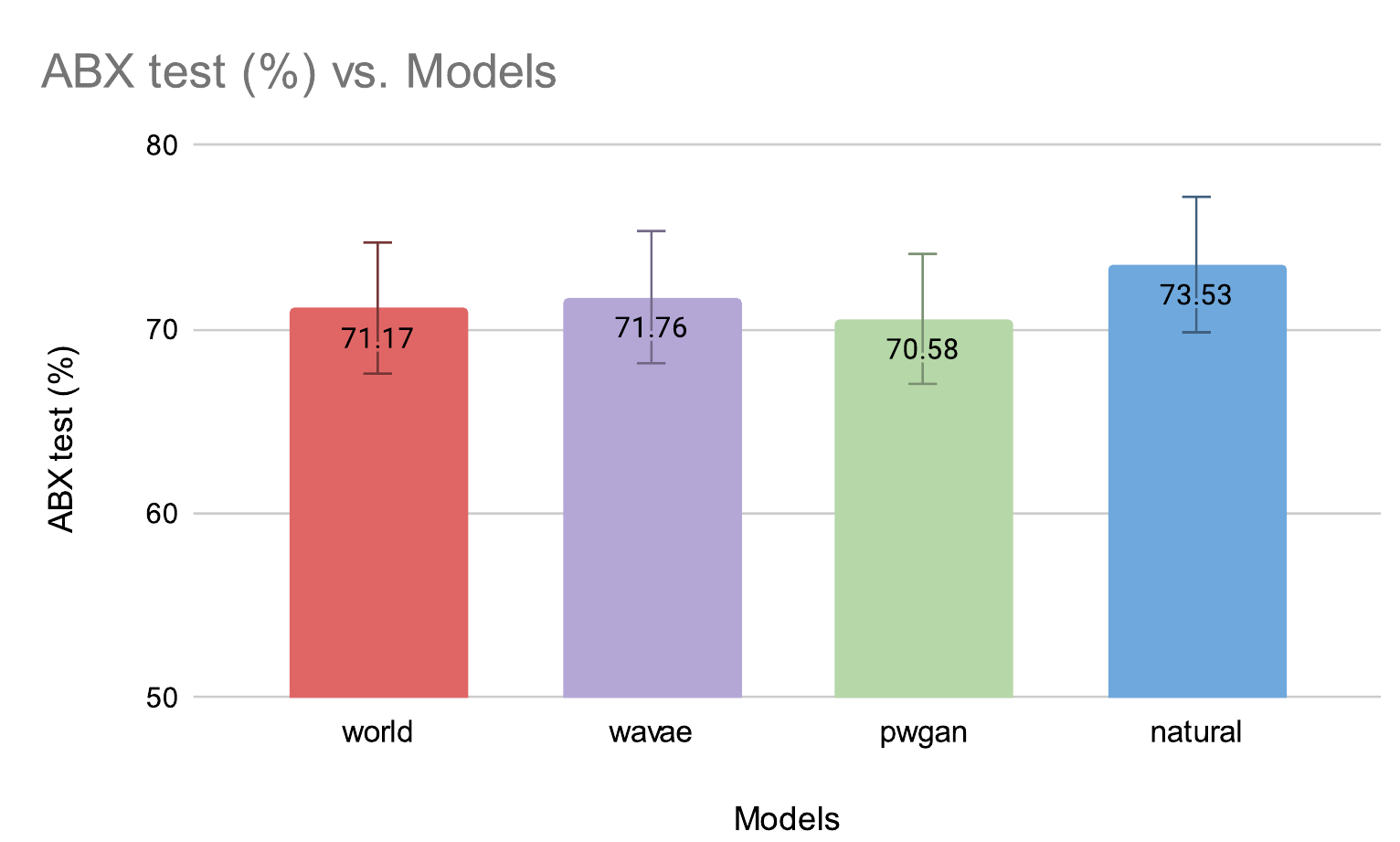}
  \caption{Summary of ABX (species recognition) results.}
  \label{fig:accuracy-summary}
\end{figure}

The ABX results summarized in Fig. \ref{fig:accuracy-summary} indicate the following:
    \begin{itemize}
        \item results for all the four cases exceed 50\%;
        \item there are no significant differences between the three vocoders, or between natural vs. vocoded samples.
    \end{itemize}
The first result suggests that naive listeners are able to identify X more accurately than by guessing, on average. The average values suggest that natural samples might be classified more accurately (73.53\%) than any of the re-synthesized samples. Due to limited sample size, however, this effect cannot be firmly confirmed.

The overall species discrimination is obviously far from perfect. There are a number of possible reasons. First, most of our subjects are non-experts who might not be exposed to birdsong on a regular basis; they may not know to which cues to pay attention to. Second, the audio samples are short. Third, originating from field recordings, some of the samples are noisy (including sounds of other birds in background). 

A look at listener-specific accuracies (not shown due to lack of space) revealed that 
the individually most accurate listener had 38 (out of 40) correct cases overall. This was the only subject who obtained perfect result on \emph{two} vocoders --- parallel WaveGAN and wavenet autoencoder. For the rest of the subjects, 
there were always at least one classification error on the vocoder samples.

\subsection{Subjective results: bird-related cues (MOS)}
\begin{figure}[t]
  \centering
  \includegraphics[width=\linewidth]{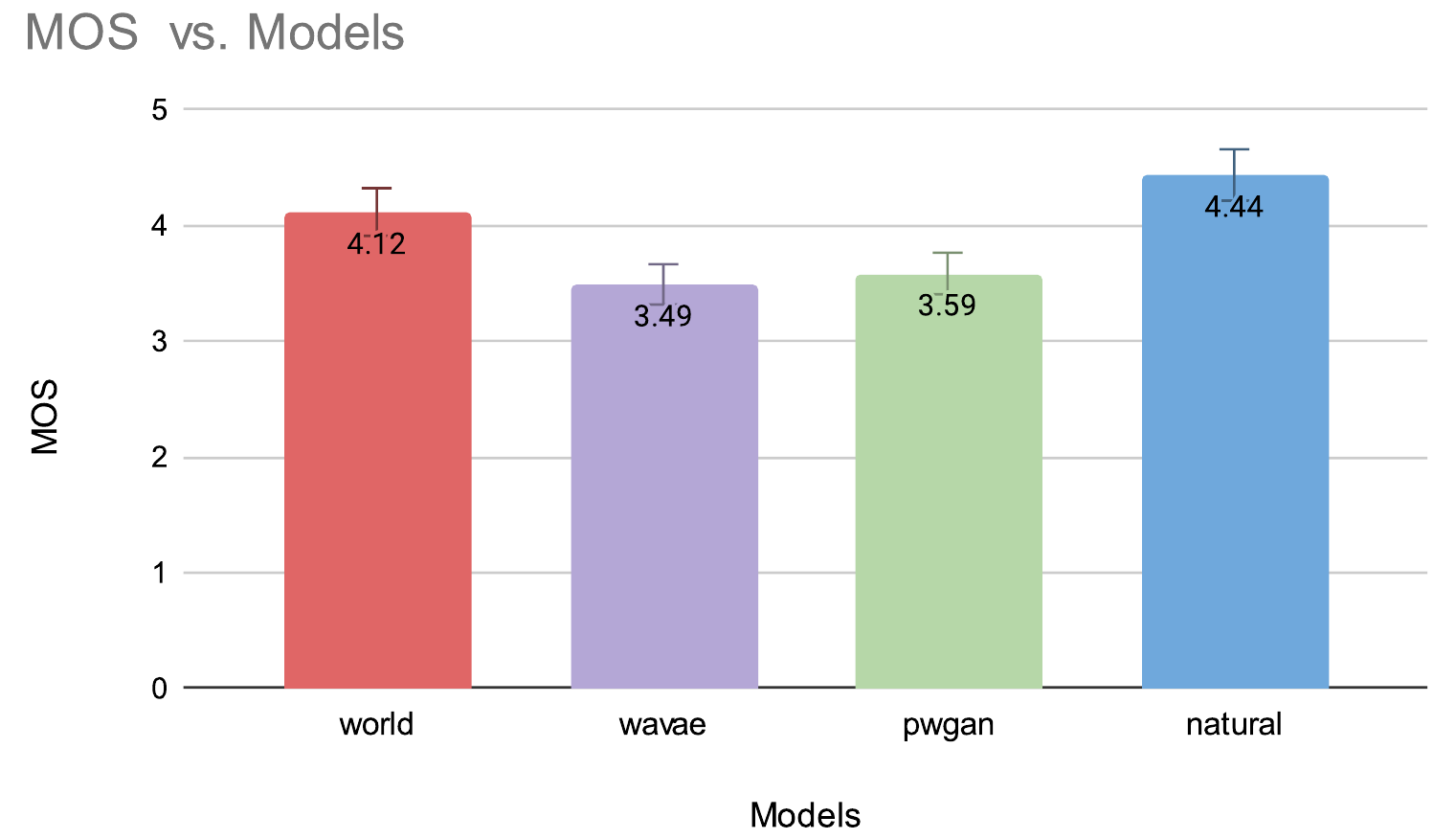}
  \caption{Summary of bird-related cue evaluation using 5-point mean opinion score (MOS) rating.}
  \label{fig:mos-summary}
\end{figure}
The results from our second experiment are summarized in Fig. \ref{fig:mos-summary}. We observe the following:
    \begin{itemize}
        \item natural samples and WORLD vocoder yield the highest MOS values with the overall range around $4.12 \sim 4.44$;
        \item the two neural vocoders yield lower MOS, WaveNet autoencoder samples being rated least `bird-like'.
    \end{itemize}

\begin{figure}[!t]
  \centering
  \includegraphics[width=\linewidth]{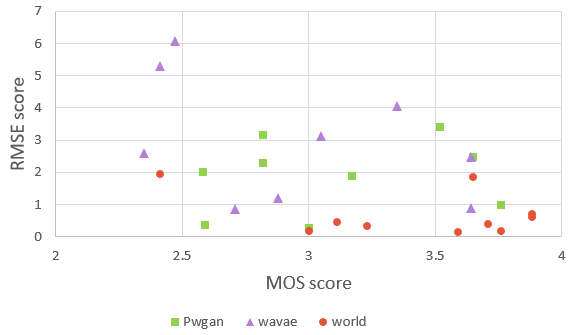}
  \caption{Scatter plot of 5-point mean opinion score (MOS) rating and root mean squared error (RMSE).}
  \label{fig:mosrmse-summary}
\end{figure}

The WORLD vocoder was deemed to retaining bird-related acoustic cues better than the two neural vocoders. The relative order of the mean responses aligns with the broad order given by the objective results (Table \ref{tab:rmse}). For completeness, RMSE vs. MOS scatterplot for individual files is displayed in Fig. \ref{fig:mosrmse-summary}. The subjective and objective measures are only weakly correlated. There is generally less RMSE variation for WORLD samples.

\section{Discussion}
\begin{figure}[!t]
  \centering
  \includegraphics[width=0.95\linewidth]{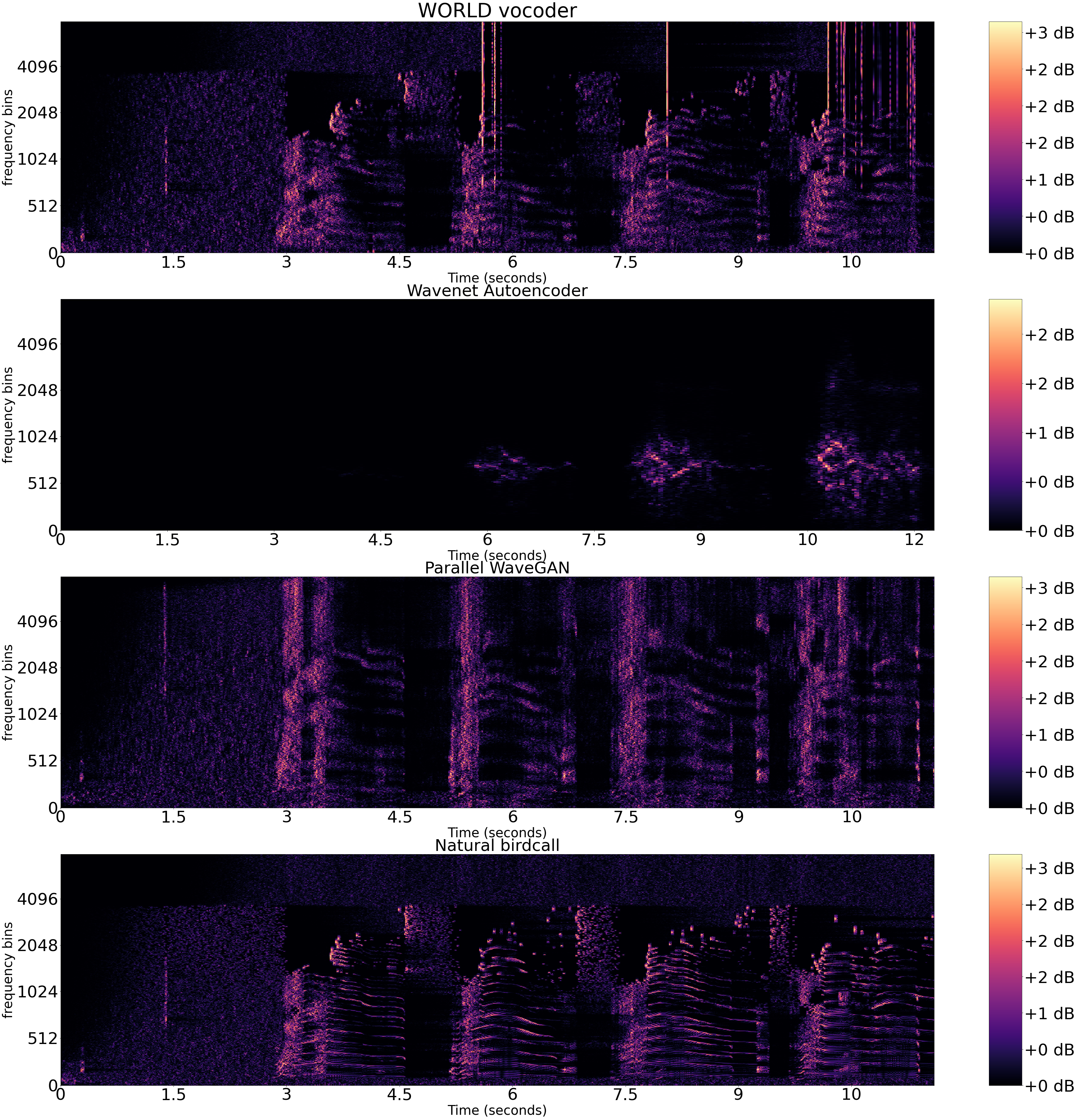}
  \caption{Natural birdcall (bottom) along with its resynthesized versions with WORLD, Wavenet AE and Parallel WaveGAN.}
  \label{fig:curious-file}
\end{figure}
Informal listening reveals strong artifacts or noisy  quality in many of the re-synthesized samples. Thus, we looked further into the individually most challenging ABX trials and lowest-rated files. For the former, we simply count the percentage of subjects who responded correctly. What constitutes a difficult trial was found to depend on the vocoder. The species pairs in the hardest trials, per vocoder, are
    \begin{itemize}
        \item Natural (60\% correct): \emph{Red-throated loon} vs. \emph{European herring gull}, \textbf{and} \emph{Eurasian blackcap} vs. \emph{Marsh warbler}
        \item WORLD (60\% correct): \emph{European goldfinch} vs. \emph{Western yellow wagtail}
        \item WaveNet Autoencoder (53\% correct): \emph{European goldfinch} vs. \emph{Western yellow wagtail}
        \item Parallel WaveGAN (53\% correct): \emph{Eurasian reed warbler} vs. \emph{European golden plover}, \textbf{and} \emph{European goldfinch} vs \emph{Western yellow wagtail}
    \end{itemize}
One particular case ---  \emph{European goldfinch} vs \emph{western yellow wagtail} --- is shared across the three vocoders. The authors believe that the difficulty does not actually relate to the species in question but to general audio properties: one of the reference samples (A or B) in these trials has very little active bird audio (but instead contains a loud splash of water, which may confuse listeners). Additionally, for all the vocoders in their respective hardest trials we noted systematic problems in reproducing voicing: the vocoders appeared to replace clear F0 trajectories in the original audio with broadband noise. 

Concerning the species pairs in \emph{easiest} cases per vocoder, two are common to all the vocoders (accuracy $\ge$ 80\% correct) ---   \emph{European nightjar} vs \emph{redwing}, \textbf{and} \emph{common moorhen} vs \emph{willow tit}. 
These cases might be easy because there is a continuous birdsong in the audio and the pairs have easily distinguishable characterstics. Additionally, some of the easy trials  contain common species that might be  \emph{familiar} even to naive listeners. For instance, $86.7 \%$ on all trials containing \emph{northern raven} were responded correctly.

As for MOS, the lowest-rated (least bird-like) sample is curiously the same for all the methods (including natural sample). The species in the said sample is \emph{red-throated loon}, with a peculiar finding: it sounds there are \emph{two} birds singing in unison but with different pitch values. The spectrograms are displayed in Fig. \ref{fig:curious-file}. For the birdsong generated by wavenet autoencoder there is silence for approximately initial $3$ seconds. Comparing each of the vocoders with the natural vocalization, WaveNet AE fails to generate harmonics. Even if WORLD and parallel WaveGAN may look better on spectrograms, listening reveals noisy and inharmonic characteristics.

\section{Conclusions}

We presented an initial study on bird vocalization resynthesis using neural vocoders. The subjective evaluation --- the main part of our study --- reveals that cues relevant for species identification are retained similarly by the three vocoders. The traditional WORLD vocoder samples, however, were rated as retaining bird-like qualities better than the two neural approaches. This finding should \emph{not} be categorically interpreted to suggest one should adopt non-neural vocoders. Since our study includes only three vocoders and one dataset, the findings should indeed be considered `initial' as suggested by the title. Future work should therefore expand the battery of vocoders and datasets.

It should be emphasized that our aim was \emph{not} to optimize a state-of-the-art birdsong resynthesis results but, rather, to carry out an out-of-the-box evaluation of existing vocoder models. Apart from retraining the deep models on bird audio, we did not consider architectural modifications, data cleaning or control parameter tuning. We expect better results to be obtained with higher-quality --- and larger amounts of --- training data. We also expect noise suppression pre-processing and different treatment of bird-present and bird-absent regions to be helpful. Overall, synthesis of acoustic wildlife data represents new challenges that are lacking from clean datasets used typically in speech synthesis and voice conversion research. The above data-engineering related questions requires detailed investigation.

For the most part, the resynthesized samples retain the important audio qualities. But it is also apparent that in a number of cases the selected vocoders struggled in reproducing correct pitch and voicing information. Apart from the challenges related to processing technically low-quality audio data, it must be kept in mind that birdsong is not the same as human voice. Thus, rather than enforcing existing \emph{voice} coders to handle bird vocalizations, in the long term it might be more beneficial to address combinations of physics-inspired and learnable models.

\section{Acknowledgment}
The work was partially sponsored by Academy of Finland. The authors express their gratitude to the subjects and to Dr. Rosa Gonz\'alez Hautamäki for her help with the PHP-forms.

\bibliographystyle{IEEEtran}
\bibliography{mybib}

\begin{thebibliography}{10}
\providecommand{\url}[1]{#1}
\csname url@samestyle\endcsname
\providecommand{\newblock}{\relax}
\providecommand{\bibinfo}[2]{#2}
\providecommand{\BIBentrySTDinterwordspacing}{\spaceskip=0pt\relax}
\providecommand{\BIBentryALTinterwordstretchfactor}{4}
\providecommand{\BIBentryALTinterwordspacing}{\spaceskip=\fontdimen2\font plus
\BIBentryALTinterwordstretchfactor\fontdimen3\font minus
  \fontdimen4\font\relax}
\providecommand{\BIBforeignlanguage}[2]{{%
\expandafter\ifx\csname l@#1\endcsname\relax
\typeout{** WARNING: IEEEtran.bst: No hyphenation pattern has been}%
\typeout{** loaded for the language `#1'. Using the pattern for}%
\typeout{** the default language instead.}%
\else
\language=\csname l@#1\endcsname
\fi
#2}}
\providecommand{\BIBdecl}{\relax}
\BIBdecl

\bibitem{Somervuo2006-parametric-bird}
\BIBentryALTinterwordspacing
P.~Somervuo, A.~H{\"{a}}rm{\"{a}}, and S.~Fagerlund, ``Parametric
  representations of bird sounds for automatic species recognition,''
  \emph{{IEEE} Trans. Speech Audio Process.}, vol.~14, no.~6, pp. 2252--2263,
  2006. [Online]. Available: \url{https://doi.org/10.1109/TASL.2006.872624}
\BIBentrySTDinterwordspacing

\bibitem{Stowell2016-bird}
D.~{Stowell}, M.~{Wood}, Y.~{Stylianou}, and H.~{Glotin}, ``Bird detection in
  audio: A survey and a challenge,'' in \emph{IEEE Int. Workshop on MLSP},
  2016, pp. 1--6.

\bibitem{Moore16-mammalian-synthesizer}
\BIBentryALTinterwordspacing
R.~K. Moore, ``A real-time parametric general-purpose mammalian vocal
  synthesiser,'' in \emph{Interspeech 2016}.\hskip 1em plus 0.5em minus
  0.4em\relax {ISCA}, 2016, pp. 2636--2640. [Online]. Available:
  \url{https://doi.org/10.21437/Interspeech.2016-841}
\BIBentrySTDinterwordspacing

\bibitem{Amador2021-synthetic}
\BIBentryALTinterwordspacing
A.~Amador and G.~B. Mindlin, ``Synthetic birdsongs as a tool to induce, and
  iisten to, replay activity in sleeping birds,'' \emph{Frontiers in
  Neuroscience}, vol.~15, p. 835, 2021. [Online]. Available:
  \url{https://www.frontiersin.org/article/10.3389/fnins.2021.647978}
\BIBentrySTDinterwordspacing

\bibitem{OReilly2016-YIN-Bird}
\BIBentryALTinterwordspacing
C.~O'Reilly, N.~M. Marples, D.~J. Kelly, and N.~Harte, ``{YIN}-bird: Improved
  pitch tracking for bird vocalisations,'' in \emph{Interspeech}.\hskip 1em
  plus 0.5em minus 0.4em\relax {ISCA}, 2016, pp. 2641--2645. [Online].
  Available: \url{https://doi.org/10.21437/Interspeech.2016-90}
\BIBentrySTDinterwordspacing

\bibitem{Bonada2016-HMM-synthesis}
\BIBentryALTinterwordspacing
J.~Bonada, R.~Lachlan, and M.~Blaauw, ``Bird song synthesis based on hidden
  {M}arkov models,'' in \emph{Interspeech 2016}, 2016, pp. 2582--2586.
  [Online]. Available: \url{http://dx.doi.org/10.21437/Interspeech.2016-1110}
\BIBentrySTDinterwordspacing

\bibitem{Gutscher2019-budgerigar}
\BIBentryALTinterwordspacing
L.~Gutscher, M.~Pucher, C.~Lozo, M.~Hoeschele, and D.~{C. Mann}, ``{Statistical
  parametric synthesis of budgerigar songs},'' in \emph{Proc. 10th ISCA Speech
  Synthesis Workshop}, 2019, pp. 127--131. [Online]. Available:
  \url{http://dx.doi.org/10.21437/SSW.2019-23}
\BIBentrySTDinterwordspacing

\bibitem{dunbar2019zero}
E.~Dunbar, R.~Algayres, J.~Karadayi, Bernard \emph{et~al.}, ``The zero resource
  speech challenge 2019: {TTS} without {T},'' \emph{arXiv preprint
  arXiv:1904.11469}, 2019.

\bibitem{tjandra2019vqvae}
A.~Tjandra, B.~Sisman, M.~Zhang, S.~Sakti, H.~Li, and S.~Nakamura, ``{VQVAE}
  unsupervised unit discovery and multi-scale code2spec inverter for zerospeech
  challenge 2019,'' \emph{arXiv preprint arXiv:1905.11449}, 2019.

\bibitem{VandenOord2016-wavenet}
A.~van~den Oord, S.~Dieleman, H.~Zen, K.~Simonyan, O.~Vinyals, A.~Graves,
  N.~Kalchbrenner, A.~W. Senior, and K.~Kavukcuoglu, ``Wavenet: {A} generative
  model for raw audio,'' in \emph{The 9th {ISCA} Speech Synthesis Workshop},
  Sunnyvale, CA, USA, September.

\bibitem{Morise2016-WORLD}
\BIBentryALTinterwordspacing
M.~Morise, F.~Yokomori, and K.~Ozawa, ``{WORLD:} {A} vocoder-based high-quality
  speech synthesis system for real-time applications,'' \emph{{IEICE} Trans.
  Inf. Syst.}, vol. 99-D, no.~7, pp. 1877--1884, 2016. [Online]. Available:
  \url{https://doi.org/10.1587/transinf.2015EDP7457}
\BIBentrySTDinterwordspacing

\bibitem{engel2017neural}
J.~Engel, C.~Resnick, A.~Roberts, Dieleman \emph{et~al.}, ``Neural audio
  synthesis of musical notes with wavenet autoencoders,'' in
  \emph{International Conference on Machine Learning}.\hskip 1em plus 0.5em
  minus 0.4em\relax PMLR, 2017, pp. 1068--1077.

\bibitem{yamamoto2020parallel}
R.~Yamamoto, E.~Song, and J.-M. Kim, ``Parallel {WaveGAN}: A fast waveform
  generation model based on generative adversarial networks with
  multi-resolution spectrogram,'' in \emph{Proc. IEEE ICASSP}, 2020, pp.
  6199--6203.

\bibitem{Kawahara1999-STRAIGHT}
\BIBentryALTinterwordspacing
H.~Kawahara, I.~Masuda{-}Katsuse, and A.~de~Cheveign{\'{e}}, ``Restructuring
  speech representations using a pitch-adaptive time-frequency smoothing and an
  instantaneous-frequency-based {F0} extraction: Possible role of a repetitive
  structure in sounds,'' \emph{Speech Commun.}, vol.~27, no. 3-4, pp. 187--207,
  1999. [Online]. Available:
  \url{https://doi.org/10.1016/S0167-6393(98)00085-5}
\BIBentrySTDinterwordspacing

\bibitem{Kawahara2008-tandem-STRAIGHT}
H.~{Kawahara}, M.~{Morise}, T.~{Takahashi}, R.~{Nisimura}, T.~{Irino}, and
  H.~{Banno}, ``Tandem-{STRAIGHT}: A temporally stable power spectral
  representation for periodic signals and applications to interference-free
  spectrum, f0, and aperiodicity estimation,'' in \emph{Proc. IEEE ICASSP},
  2008, pp. 3933--3936.

\bibitem{morise2016d4c}
M.~Morise, ``{D4C}, a band-aperiodicity estimator for high-quality speech
  synthesis,'' \emph{Speech Communication}, vol.~84, pp. 57--65, 2016.

\bibitem{Goodfellow2014-GAN}
\BIBentryALTinterwordspacing
I.~J. Goodfellow, J.~Pouget{-}Abadie, M.~Mirza, B.~Xu, D.~Warde{-}Farley,
  S.~Ozair, A.~C. Courville, and Y.~Bengio, ``Generative adversarial nets,'' in
  \emph{Proc. {NIPS}}, 2014, pp. 2672--2680. [Online]. Available:
  \url{https://proceedings.neurips.cc/paper/2014/hash/5ca3e9b122f61f8f06494c97b1afccf3-Abstract.html}
\BIBentrySTDinterwordspacing

\bibitem{salimans2016weight}
T.~Salimans and D.~P. Kingma, ``Weight normalization: A simple
  reparameterization to accelerate training of deep neural networks,''
  \emph{arXiv preprint arXiv:1602.07868}, 2016.

\bibitem{stowell2014automatic}
D.~Stowell and M.~D. Plumbley, ``Automatic large-scale classification of bird
  sounds is strongly improved by unsupervised feature learning,'' \emph{PeerJ},
  vol.~2, p. e488, 2014.

\bibitem{web:xenocanto}
\BIBentryALTinterwordspacing
``xeno-canto --- sharing bird sounds from around the world,'' 2017, last
  accessed 11 Marc 2021. [Online]. Available: \url{https://www.xeno-canto.org/}
\BIBentrySTDinterwordspacing

\bibitem{haque2018conditional}
A.~Haque, M.~Guo, and P.~Verma, ``Conditional end-to-end audio transforms,''
  \emph{arXiv preprint arXiv:1804.00047}, 2018.

\bibitem{imai2009speech}
S.~Imai, T.~Kobayashi, K.~Tokuda, T.~Masuko, K.~Koishida, S.~Sako, and H.~Zen,
  ``Speech signal processing toolkit (sptk),'' 2009.

\bibitem{web:ffmpeg}
\BIBentryALTinterwordspacing
W.~Robitza, ``ffmpeg tool,'' 2015, last accessed 11 March 2021. [Online].
  Available: \url{https://github.com/slhck/ffmpeg-normalize}
\BIBentrySTDinterwordspacing

\end{thebibliography}

\end{document}